\documentclass[graybox]{svmult}

\pdfoutput=1

\usepackage{mathptmx}       
\usepackage{helvet}         
\usepackage{courier}        
\usepackage{type1cm}        
\usepackage{makeidx}         
\usepackage{graphicx}        
\usepackage{multicol}        
\usepackage[bottom]{footmisc}

\usepackage{amsmath}
\usepackage{amsfonts}

\usepackage{tikz}
\usetikzlibrary{patterns,decorations.pathreplacing}
\usetikzlibrary{arrows.meta}
\usetikzlibrary{shapes.arrows, fadings}

\let\originalleft\left
\let\originalright\right
\renewcommand{\left}{\mathopen{}\mathclose\bgroup\originalleft}
\renewcommand{\right}{\aftergroup\egroup\originalright}

\providecommand{\df}{\textrm{d}}

\newcommand{\pdiff}[3][]{\frac{\partial^{#1}#2}{\partial{#3}^{#1}}}
\newcommand{\Es}{E_{\textrm{sat}}}
\newcommand{\FT}[1]{\mathcal{F}\left\{ #1 \right\}}
\newcommand{\FTi}[1]{\mathcal{F}^{-1}\left\{ #1 \right\}}

\makeindex             

\begin{document}

\title*{A New Method of Modelling Tuneable Lasers with Functional Composition}
\author{B. Metherall \and C. S. Bohun}
\institute{B. Metherall \at University of Oxford,  Oxford, UK, \email{brady.metherall@maths.ox.ac.uk}
\and C. S. Bohun \at Ontario Tech University, 2000 Simcoe St N, Oshawa, ON, \email{sean.bohun@uoit.ca}}
%
%
\maketitle

\abstract*{
A new nonlinear model is proposed for tuneable lasers. Using the generalized nonlinear Schr\"{o}dinger equation as a starting point, expressions for the transformations undergone by the pulse are derived for each of the five components (gain, loss, dispersion, modulation, and nonlinearity) within the laser cavity. These transformations are then composed to give the overall effect of one trip around the cavity. We examine the full nonlinear model which is solved numerically. A consequence of the nonlinear nature of this model is that it is able to exhibit wave breaking which prior models could not. We highlight the rich structure of the boundary of stability for a particular plane of the parameter space.
}

\abstract{
A new nonlinear model is proposed for tuneable lasers. Using the generalized nonlinear Schr\"{o}dinger equation as a starting point, expressions for the transformations undergone by the pulse are derived for each of the five components (gain, loss, dispersion, modulation, and nonlinearity) within the laser cavity. These transformations are then composed to give the overall effect of one trip around the cavity.
}

\section{Introduction}
\label{sec:intro}
\begin{figure}[tbp]
\centering

\begin{tikzpicture}
\draw [rounded corners=4mm] (0,0) rectangle ++(6,4);
\draw [rounded corners=4mm] (0,0) rectangle ++(-1.5,4);

\draw (0.5,2.25) circle (0.5cm);
\draw (0.5,2) circle (0.5cm) node [anchor=west,xshift=0.5cm,align=center] {Er-Doped \\ Fibre};
\draw (0.5,1.75) circle (0.5cm);

\filldraw[fill=white, draw=black] (2,-0.75) rectangle ++(2,1.5) node [midway] {Modulator};
\filldraw[fill=white, draw=black] (-2.1,1.5) rectangle ++(1.2,1) node [midway] {Pump};

\draw[-stealth] (3,4) -- (3,5.5) node [pos=0.75,anchor=west,xshift=0.25cm] {Laser Output};
\draw[densely dashdotted] (2.5,3.5) -- (3.5,4.5) node [pos=1,anchor=north,yshift=-0.75cm,align=center] {Optical \\ Coupler};

\filldraw[fill=white, draw=black] (6,2) circle (0.5cm) node [anchor=west,xshift=0.5cm,align=center] {Optical \\ Circulator};
\draw[->,>=stealth] (6,2.325) arc (90:360:0.325cm);

\filldraw[fill=white, draw=black] (0,0) circle (0.5cm);
\draw[->,>=stealth] (0,0.325) arc (90:360:0.325cm);

\filldraw[fill=white, draw=black] (0,4) circle (0.5cm);
\draw[->,>=stealth] (0,4.325) arc (90:360:0.325cm);

\draw [->,>=stealth,domain=20:70] plot ({0.675*cos(\x)}, {0.675*sin(\x)});
\draw [->,>=stealth,domain=110:160] plot ({0.675*cos(\x)}, {0.675*sin(\x)});
\draw [->,>=stealth,domain=110:160] plot ({6+0.675*cos(\x)}, {2+0.675*sin(\x)});
\draw [->,>=stealth,domain=200:250] plot ({6+0.675*cos(\x)}, {2+0.675*sin(\x)});
\draw [->,>=stealth,domain=200:250] plot ({0.675*cos(\x)}, {4+0.675*sin(\x)});
\draw [->,>=stealth,domain=290:340] plot ({0.675*cos(\x)}, {4+0.675*sin(\x)});

\draw (5.5,2) -- (3.5,2) node [pos=0.5,anchor=south,yshift=0.25cm,xshift=-0.15cm] {CFBG};
\foreach \i in {0,...,13}
  \draw (3.5 + \i*\i/100,1.75) -- (3.5 + \i*\i/100,2.25);

\end{tikzpicture}
\caption{Typical cavity of a fibre based tuneable laser. The laser pulses travel clockwise around each loop.}
\label{fig:cavity}
\end{figure}
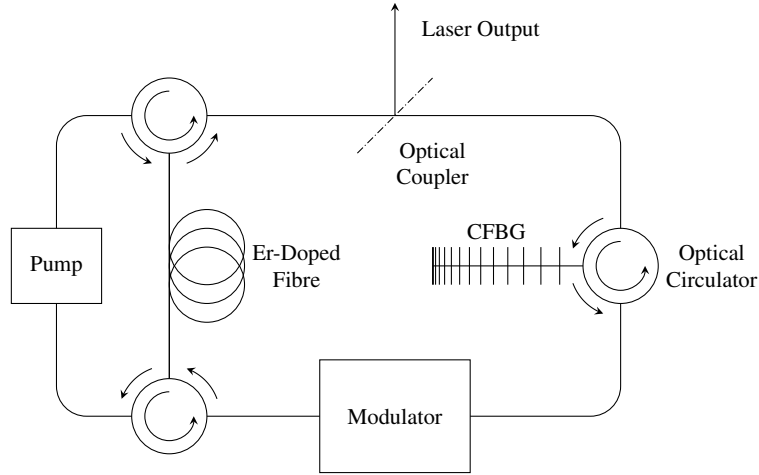

A tuneable laser has the ability to vary the frequency of its output by up to about 100 nanometres \cite{bohun, burgoyne2010, yamashita}. Tuneable lasers simultaneously lase at all frequencies within this bandwidth. This tuneability is quite useful and has applications in spectroscopy and high resolution imaging such as coherent anti-Stokes Raman spectroscopy and optical coherence tomography \cite{bohun, burgoyne2014, yamashita}, as well as communications and diagnostics of ultra fast processes \cite{silfvast}. A typical tuneable laser cavity is shown in Figure \ref{fig:cavity}. In contrast to a standard laser, a tuneable laser contains two additional components, namely, a chirped fibre Bragg grating (CFBG), and a modulator.

A CFBG is a length of optical fibre where the refractive index oscillates along its length \cite{ferreira}, and therefore, can act as a reflective filter \cite{agrawal2002, alazzawi, ferreira, starodoumov}. Due to the oscillatory nature, light with the corresponding wavelength will be reflected when the Bragg condition is satisfied \cite{agrawal2002, alazzawi, becker, ferreira, silfvast, starodoumov}. The spacial variation of the refractive index effectively creates a spacial dependence on the Bragg condition, causing most wavelengths to be reflected by a CFBG, but with each wavelength satisfying the Bragg condition at a different spacial location\footnote{Note that a monotonic chirping ensures that the spacial dependence of the Bragg condition is continuous with respect to the frequency.}. A consequence of this is that a time delay is created between wavelengths---this causes the pulse to disperse and broaden.

The modulator serves the purpose of reshaping the pulse. Without it, the pulse will repeatedly widen due to the CFBG---the modulator ensures the pulse is band limited by altering the envelope. 

\section{Modelling Efforts}
\label{sec:modelling}
The standard equation for studying nonlinear optics is the nonlinear Schr\"odinger equation (NLSE)\footnote{The NLSE can be derived from the nonlinear wave equation for electric fields; this derivation is presented in detail in \cite{agrawal2013, ferreira}.},
\begin{align}
\pdiff{A}{z} &= - i \frac{\beta_2}{2}\pdiff[2]{A}{T} + i \gamma |A|^2 A.
\label{eq:smallnlse}
\end{align}
Here $A = A(T, z) : \mathbb{R}^2 \mapsto \mathbb{C}$ is the complex pulse amplitude, $\beta_2 \in \mathbb{R}$ is the second order dispersion, and $\gamma \in \mathbb{R}$ is the coefficient of nonlinearity. In practice, \eqref{eq:smallnlse} lacks a few key terms, thus, it is often generalized by adding amplification, loss, and occasionally higher order terms. This gives the generalized nonlinear Schr\"{o}dinger equation (GNLSE) \cite{agrawal2013, bohun, finot, peng, shtyrina, yarutkina},
\begin{align}
\label{eq:nlse}
\pdiff{A}{z} &= - i \frac{\beta_2}{2}\pdiff[2]{A}{T} + i \gamma |A|^2 A + \frac{1}{2}g(A) A - \alpha A,
\end{align}
where $g(A)$ is an amplifying term due to the gain, and $\alpha \in \mathbb{R}$ is the loss due to scattering and absorption.

The GNLSE has many applications in nonlinear optics and fibre optic communications, however, in the context of lasers we typically also add a modulation term. This yields the master equation of mode-locking \cite{haus1975, hausbook, haus1986, haus1992, haus2000, kartner, tamura1996, usechak},
\begin{align}
\pdiff{A}{z} &= - i \frac{\beta_2}{2}\pdiff[2]{A}{T} + i \gamma |A|^2 A + \frac{1}{2}g(A) A - \alpha A - M(T),
\label{eq:meml}
\end{align}
where $M(T)$ is the modulation function. The solutions of three simplifications of \eqref{eq:meml} have been investigated:
\begin{itemize}
\item Omitting both modulation and nonlinearity \cite{haus1975, haus1986, haus1992}.
\item Omitting only modulation \cite{haus1991, usechak}.
\item Omitting only nonlinearity  \cite{burgoyne2014, haus1975, hausbook, haus1996, haus2000, kartner, tamura1996, usechak}.
\end{itemize}
For a more comprehensive history see \cite{haus2000}.

\subsection{Discrete Models}
While the derivation of \eqref{eq:meml} is correct mathematically, it is not representative of what happens within the laser cavity. The issue with \eqref{eq:meml} is that it has been assumed each process affects the pulse continuously within the cavity. As highlighted by Figure \ref{fig:cavity}, this is a rather poor assumption. Within the cavity each effect is localized to its corresponding component: almost all of the dispersion happens within the CFBG \cite{agrawal2002}, the pulse is only amplified within the Erbium-doped fibre, etc. Thus, perhaps a better model is one where \eqref{eq:meml} is broken down into the individual components giving the effect of each `block' of the cavity. Each of the blocks can then be functionally composed together to give an iterative map for the effect of one circuit around the cavity. This transforms the differential equation into an algebraic equation.

Such a method was first proposed in 1955 by Cutler \cite{cutler} while analyzing a microwave regenerative pulse generator. This method was adapted for mode-locked lasers in 1969 by Siegman and Kuizenga \cite{kuizenga1970a, siegman}. Kuizenga and Siegman also had success experimentally validating their model \cite{kuizenga1970b, kuizenga1970}. The effects of the nonlinearity would not be considered until Martinez, Fork, and Gordon \cite{martinez1984, martinez1985} tried modelling passively mode-locked lasers. This issue has recently been readdressed by Burgoyne \cite{burgoyne2014} in the literature for tuneable lasers. In these models the effect of each block is described by a transfer function.

Despite the development of these block style models, several short-comings exist. The clearest is that none of these models have contained every block---either the nonlinearity or the modulation have been omitted. In the framework of tuneable lasers, each component plays a crucial role and the tuneable laser will not function correctly without the inclusion of all of the components. Another key drawback is that the functional operations of some of the components used in their models are phenomenological. While these functions are chosen based on the observed output, they are not necessarily consistent with the underlying physics. Finally, none of these previous models have been able to exhibit a phenomena called \emph{wave-breaking} in which the self-phase modulation of the pulse becomes too strong, distorting and damaging the wave until it ultimately becomes unstable and unsustainable.

\section{A New Model}
\label{sec:newmodel}
Using the ideas presented in the previous section of the prior functional models \cite{burgoyne2014, cutler, kuizenga1970a, kuizenga1970b, kuizenga1970, martinez1984, martinez1985, siegman} we shall derive a new model from \eqref{eq:nlse}---with the exception of the modulation in which we consider the exact functional form to be determined by the laser operator.

\subsection{Components}
We shall determine the effect each component has on the pulse by solving \eqref{eq:nlse} while only considering the dominant term within each section of the cavity.

\subsubsection{Gain}
Within the Er-doped gain fibre, the gain term is dominant, and equation \eqref{eq:nlse} reduces to
\begin{align}
\label{eq:gainde}
\pdiff{A}{z} &= \frac{1}{2} g(A) A,
\end{align}
where $g(A)$ takes the form \cite{bohun, burgoyne2014, haus1975, hausbook, haus1992, haus2000, haus1991, kartner, peng, shtyrina, silfvast, usechak, yarutkina}
\begin{align}
\label{eq:energy}
	g(A) &= \frac{g_0}{1 + E / \Es},& E &= \int_{-\infty}^\infty |A|^2 \, \df T,
\end{align}
where $g_0$ is a small signal gain, $E$ is the energy of the pulse, and $\Es$ is the energy at which the gain begins to saturate. Without much difficulty this differential equation can be solved, and the effect on an incident pulse is
\begin{align}
G(A;E) &= \left(\frac{E_{\textrm{out}}}{E}\right)^{1/2}A = \left( \frac{\Es}{E} W \left( \frac{E}{\Es} \textrm{e}^{E/\Es} \textrm{e}^{g_0 L_g} \right) \right)^{1/2} A,
\end{align}
where $L_g$ is the length of the gain fibre.

\subsubsection{Nonlinearity}
The nonlinearity of the fibre arises from the parameter $\gamma$; in regions where this effect is dominant expression~\eqref{eq:nlse} becomes
\begin{align}
\label{eq:fibrediff}
	\pdiff{A}{z} - i \gamma |A|^2 A = 0.
\end{align}
Using a similar method as with the gain, the effect of the nonlinearity can be shown to be
\begin{align}
F(A) &= A \textrm{e}^{i \gamma |A|^2 L_f},
\end{align}
where $L_f$ is the length of fibre.

\subsubsection{Loss}
Expression \eqref{eq:nlse} leads to exponential decay due to the scattering and absorption of the fibre. However, a majority of the signal is removed from the cavity by the optical coupler. Combining these two effects yields
\begin{align}
L(A) &= (1 - R) \textrm{e}^{- \alpha L_T}A,
\end{align}
where $R$ is the reflectivity of the output coupler, and $L_T$ is the total length of the laser circuit as the effect of the losses\footnote{Depending on the layout of the laser cavity the loss may take the form $L(A) = R \textrm{e}^{- \alpha L_T}A$ instead.}.

\subsubsection{Dispersion}
Considering only the dispersive terms of \eqref{eq:nlse}, one obtains
\begin{align}
\label{eq:dispde}
	\pdiff{A}{z} = -i \frac{\beta_2}{2} \pdiff[2]{A}{T}.
\end{align}
The effect of dispersion is then given by the map
\begin{align}
D(A) &= \FTi{\textrm{e}^{i \omega^2 L_D\beta_2/2} \FT{A}},
\end{align}
where $L_D$ is the length of the dispersive medium, and $\mathcal{F}$ denotes the Fourier transform.

\subsubsection{Modulation}
In this model, the modulation is considered to be applied externally in which ever way the operator sees fit. For simplicity the representation is taken as the Gaussian
\begin{align}
M(A) &= \textrm{e}^{-T^2 / 2 T_M^2} A,
\end{align}
where $T_M$ is the characteristic width of the modulation.

\subsection{Non-Dimensionalization}
The structure of each process of the laser can be better understood by re-scaling the time, energy, and amplitude. Nominal values for tuneable lasers are shown in Table \ref{tab:values}. Knowing experimental durations and energies, the table suggests the convenient scalings:
\begin{align}
	T &= T_M \widetilde{T},& E &= \Es \widetilde{E},& A &= \left( \frac{\Es}{T_M} \right)^{1/2} \widetilde{A}.
\end{align}
Revisiting each process map shows each process has a characteristic non-dimensional parameter. The new mappings---after dropping the tildes---are
\begin{equation}
\begin{aligned}
G(A) &= \left(E^{-1} W \left( a E \textrm{e}^{E}\right) \right)^{1/2} A, & F(A) &= A \textrm{e}^{i b |A|^2}, & L(A) &= h A, \\
D(A) &= \FTi{\textrm{e}^{i s^2 \omega^2} \FT{A}}, & M(A) &= \textrm{e}^{-T^2 / 2} A,
\label{eq:effects}
\end{aligned}
\end{equation}
with the four dimensionless parameters, as defined by the values in Table \ref{tab:values},
\begin{equation}
\begin{aligned}
	a &= \textrm{e}^{g_0 L_g} \sim 8 \times 10^3,& \qquad h &= (1 - R) \textrm{e}^{-\alpha L} \sim 0.04, \\
	b &= \gamma L_f \frac{\Es}{T_M} \sim 1,& \qquad s &= \sqrt{\frac{\beta_2 L_D}{2 T_M^2}} \sim 0.2,
\label{eq:ndparam}
\end{aligned}
\end{equation}
which characterize the behaviour of the laser. Notice that the modulation is only characterized by $T_M$, and each other process has its own independent non-dimensional parameter.

\begin{table}[tbp]
\caption{Range of variation of various parameters.}
\label{tab:values}
\centering
\begin{tabular}{lcll}
\hline\noalign{\smallskip}
Parameter & Symbol & Value & Sources \\
\hline\noalign{\smallskip}
Absorption of Fibre$^a$& $\alpha$ & $10^{-4}$--$0.3\text{ m}^{-1}$  & \cite{burgoyneemail, shtyrina, tomlinson, usechak, yarutkina} \\
Fibre Dispersion & $\beta_2^f$ & $-50$--$50 \text{ ps}^2/ \text{km}$ & \cite{agrawal2002, agrawal2013, burgoyne2014, litchinitser, peng, yarutkina} \\
Fibre Nonlinearity & $\gamma$ & $0.001$--$0.01 \text{ W}^{-1} \text{m}^{-1}$ & \cite{agrawal2013, finot, usechak, yarutkina} \\
Grating Dispersion & $\beta_2^g L_D$ & $10$--$2000 \text{ ps}^2$ & \cite{agrawal2002, agrawal2013, burgoyne2014, li} \\
Length of Cavity & $L_T$ & $10$--$100 \text{ m}$ & \cite{burgoyneemail, peng, tamura1996} \\
Length of Fibre & $L_f$ & $0.15$--$1 \text{ m}$ & \cite{burgoyneemail} \\
Length of Gain Fibre & $L_g$ & $2$--$3 \text{ m}$ & \cite{burgoyne2014, peng, shtyrina, tamura1993, yarutkina} \\
Modulation Time & $T_M$ & $15$--$150 \text{ ps}$ & \cite{bohun, burgoyneemail, burgoyne2014} \\
Reflectivity of Optical Coupler & $R$ & $0.1$--$0.9$ & \cite{burgoyneemail, li, peng,  tamura1993, tamura1996, yamashita} \\
Saturation Energy & $\Es$ & $10^3$--$10^4 \text{ pJ}$ & \cite{burgoyneemail, usechak, yarutkina} \\
Small Signal Gain & $g_0$ & $1$--$10 \text{ m}^{-1}$ & \cite{burgoyneemail, yarutkina} \\
\noalign{\smallskip}\hline\noalign{\smallskip}
\end{tabular} \\
$^a$  Fibre loss is typically reported as $\sim0.5$ dB/km.
\end{table}

\subsection{Combining the Effects of Each Block of the Model}
\label{sec:effects}
In this model the pulse is iteratively passed through each process, the order of which must now be considered. We are most interested in the output of the laser cavity, and so we shall start with the loss component. Next the pulse is passed though the CFBG, as well as the modulator. Finally, the pulse travels through the gain fibre to be amplified, and then we consider the effect of the nonlinearity since this is the region where the power is maximal. Note that in general the functional operators of the components do not commute, and therefore the order of the components is indeed important---in contrast to the previous models. This is especially the case of dispersion as realized through the Fourier transform. Functionally this can be denoted as
\begin{align}
\mathcal{L}(A) = F(G(M(D(L(A))))),
\end{align}
where $\mathcal{L}$ denotes one loop of the laser. The pulse after one complete circuit of the laser cavity is then passed back in to restart the process. A steady solution to this model is one in which the envelope and chirp are unchanged after traversing every component in the cavity---we are uninterested in the phase. That is, such that $\mathcal{L}(A) = A \textrm{e}^{i \phi}$---for some $\phi \in \mathbb{R}$.

\section{Conclusion}
Within this paper we developed a new nonlinear model for tuneable lasers. In order to better represent the underlying physics within the laser cavity the nonlinear Schr\"{o}dinger equation was reduced to simpler differential equations for each component of the laser. This led to a functional map that defines the effect of each component on a particular input pulse. These processes were then composed together to give an iterative mapping of the whole laser cavity. In a future publication we shall show the results obtained by this iterative mapping as well as discuss the dynamics exhibited by this model---including wave-breaking---to predict the conditions under which the pulse is stable and sustainable.

\bibliography{Ref}

\end{document}